\documentclass[aps,pre,reprint,superscriptaddress,amsmath,showpacs]{revtex4-1}
\usepackage[breaklinks,colorlinks=true]{hyperref}
\usepackage{graphicx}
\usepackage{amssymb}
\usepackage{amsmath}
\usepackage{verbatim}
\usepackage{chngcntr}

\begin{document}

\title{Emergence of Long-Term Memory in Popularity}

\author{Hyungjoon Soh}
\affiliation{Department of Physics,
	Korea Advanced Institute of Science and Technology, Daejeon
	34141, Korea}

\author{Joo Hyung Hong} 
\affiliation{Department of Global Management, KyungHee Cyber University, Seoul 02447, Republic of Korea}

\author{Jaeseung Jeong}
\affiliation{Department of Bio and Brain Engineering, KAIST, Daejeon 34141, Republic of Korea}

\author{Hawoong Jeong}
\email[Corresponding author: ]{hjeong@kaist.edu}
\affiliation{Department of Physics and Institute for the
	BioCentury, Korea Advanced Institute of Science and Technology,
	Daejeon 34141, Korea}

\date{\today}

\begin{abstract}
Popularity describes the dynamics of mass attention, and is a part of a broader class of population dynamics in ecology and social science literature. Studying accurate model of popularity is important for quantifying spreading of novelty, memes, and influences in human society. Although logistic equation and similar class of nonlinear differential equation formulates traditional population dynamics well, part of the deviation in long-term prediction is stated, yet fully understood. Recently, several studies hinted a long-term memory effect on popularity whose response function follows a power-law, especially that appears on online mass media such as YouTube, Twitter, or Amazon book sales. Here, we investigate the ranking of most popular music, \textit{the Billboard Hot 100 chart}, which is one of the largest popularity dataset spanning several decades. Using a popularity model that comprises logistic growth and a power-law decaying long-term memory, we showed that rank history is mainly characterized by initial popularity and memory strength. With this framework, we investigated temporal development of long-term memory on the whole popularity dynamics. As a result, abrupt emergence of long-term memory and broad initial popularity is illustrated, which was not clearly detected by time-independent measures. We emphasize not only development of the mass media, but also the difference of spreading and accumulated popularity affect dynamics significantly when the popularity has long-term memory.
  
\end{abstract}
\maketitle

\section{Introduction}

Popularity is an important measure that quantifies impact of things around everyday life. Since there is too much information to choose items nowadays, people often rely on popularity to reduce time and cost of choice; in other words, popularity acts as a driving force to every choice. On the other side, popularity also determines any product's fate, since often its sales are thoroughly related popularity in many of its characters, \textit{e.g.}, price, quality, and so on. Moreover, studying popularity is emphasized by its nonlinear character, which makes less predictable. Dynamics of popularity is studied in many contexts, such as citation in research articles~\cite{DWang2013, YHEom2011, Mazloumian2011}, box office revenue~\cite{Ishii2012, Chakrabarti2016, Pan2010, Sinha2004}, Twitter retweets, social networks services (SNSs)~\cite{FWu2007}, YouTube video views~\cite{Crane2008, Chatzopoulou2010, Figueiredo2011, Mitchell2010}, Amazon book sales ~\cite{Sornette2004, Deschatres2005}, as well as many other items~\cite{Kawahata2013, SDKim2012, Yasseri2012}. 


Especially, the music popularity data has been most preferable, compared to others; the pressure which makes the music more popular. Many other items have ``{\em inertia}'', which is some kind of cost to consume or adapt such items - physical limitation due to finite supply, demand, or transportation. However, the music does not suffer the inertia problem since there is no external factor, but the cost itself for music consumption. Another virtue of the popularity of music is that the pressure is mostly voluntary. Some types of popularity, \textit{e.g.} paper citation, has finite lifetime so that every research draws scientific community's attention until new novelty arises. Another example would be electronic devices, where the novelty is closely related to external factors, such as hardware technology or software compatibility. However, music does not have capacity nor inertia, which guarantees the consumption itself is proportional to popularity itself.

Among all of the musical popularity measures, we studied {\em the Billboard Chart} to study mesoscopic dynamics of popularity. The Billboard chart records popularity trend of the music for more than half a century, which is long enough to discuss evolution of the way how the music is produced and consumed. Also, since the Billboard chart ranks the music with highest popularity, the popularity measure is robust to the external noise. These parts have been benefits to study the Billboard charts as a popularity measures in numerous studies~\cite{Bhattacharjee2007, Bhattacharjee2007a, Bradlow2001}.

The one focus we extensively discuss in this work is how the rank changed over time. Before the gramophone was invented, the music was consumed as soon as they were produced. Therefore, the music gained popularity with indirect impression only. After gramophone, music was able to spend as physical material. Then, the radio and broadcasting lead to global spread of the music. Nowadays, personal devices and internet streaming lead to new dynamics of popularity evolution. Not only the popularity growth speed changed, but also its distribution as well~\cite{DWang2013}.

To explain long-term trend in popularity dynamics, long-range temporal correlation which does not appear in classical Markovian population dynamics is suspected to play an important role. Particularly, long-term memory which appears in various social systems~\cite{Sornette2004, Deschatres2005} makes key difference with most of probabilistic model systems. Several methods are used to embed long-range correlation, by using extra degrees of randomness such as random matrices or model underlying noise dynamics. 

In our work, we investigate popularity dynamics and its evolution by data of long history. We first parametrize each single's rank history and look into statistics to understand the general picture of music popularity. Long-term trend of popularity dynamics that how people spend the music is investigated. To discuss long-term effect, we separate the time sections as 5 segments by important events and compare each segment statistics. 


\begin{figure}[]
 \centering
    \includegraphics[width=\columnwidth]{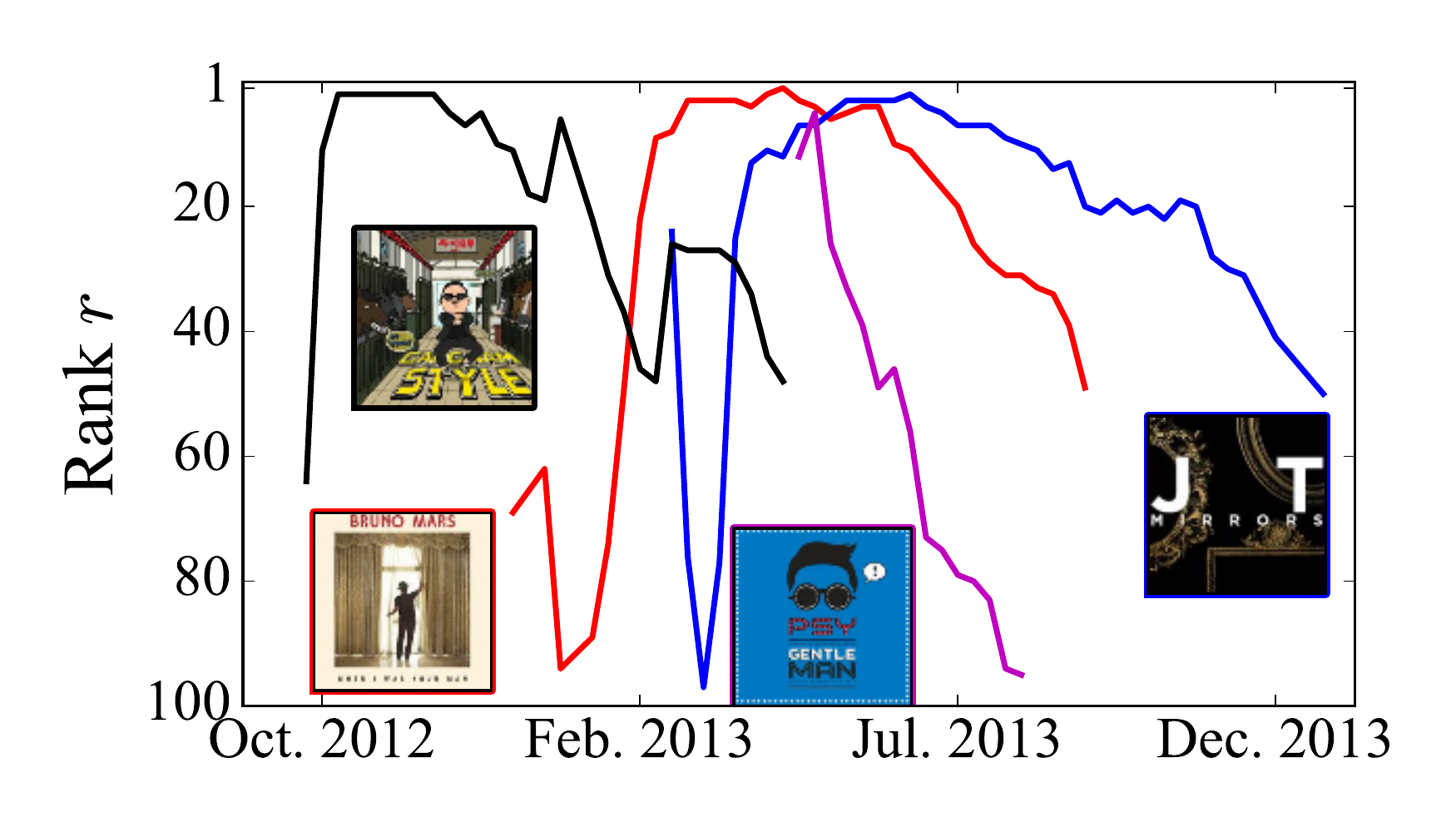}
	\caption{\label{fig:PathExp} Example of rank path in a {\em Billboard} Hot 100 chart. Four songs in October 2012 to December 2013 are presented, {\em Gangnam Style}-PSY (black), {\em Gentleman}-PSY (purple), {\em When I Was Your Man}-Bruno Mars (red), and {\em Mirrors}-Justin Timberlake (blue).}
\end{figure}

\begin{figure*}[t]
 \centering
   \includegraphics[width=17.8cm]{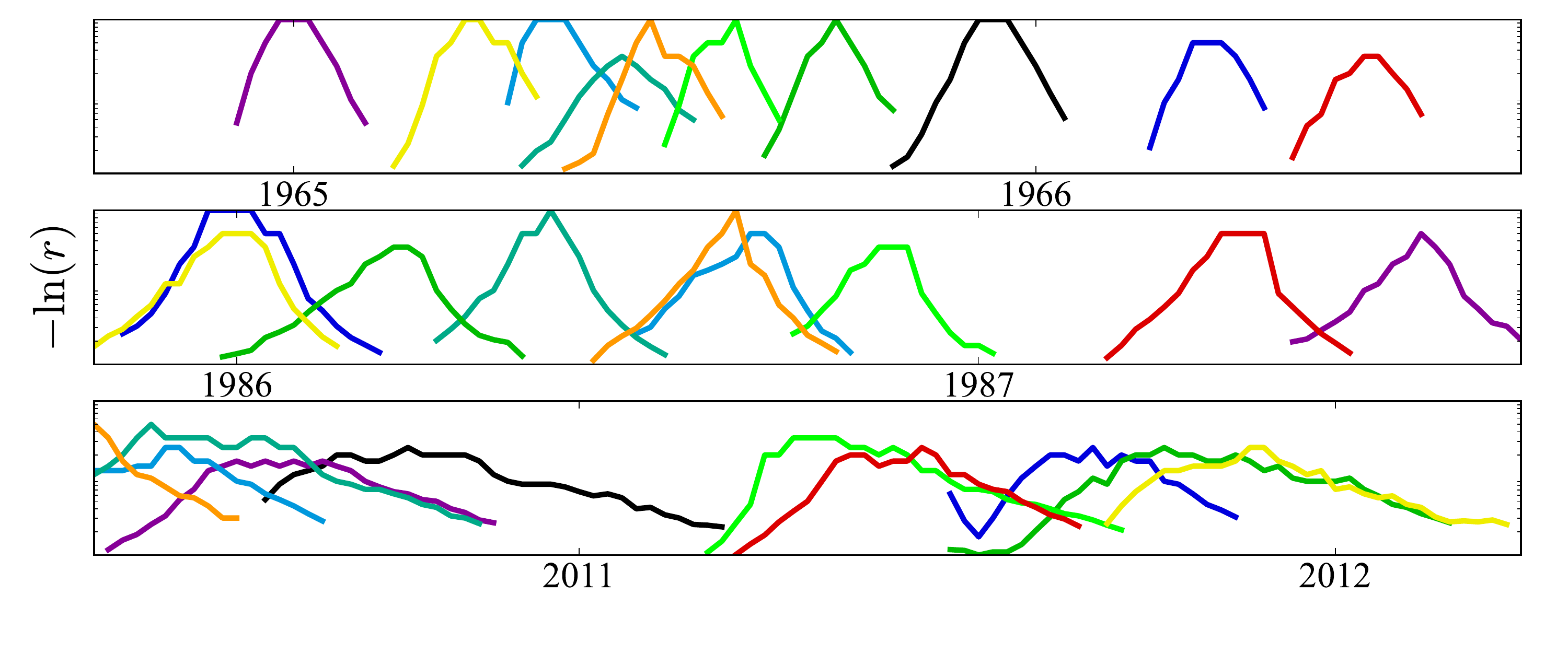}
	\caption{\label{fig:PopMap} Popularity comparison by time in early (1958-1960, upper panel), intermediate (1985-1986, middle panel), and later (2012-2014, lower panel). 10 songs with long lifetime in each time frame are shown in each panel. The value $\ln (1/r)$ is used rather than the rank itself was used for comparison.}
\end{figure*}

\section*{Results}
\subsection{Rank movement in a chart}\label{sec:BasicStats}
When a song enters into chart, its rank follows the five stage of life: Enter-Rise-Sustain-Fall-Exit (Fig.~\ref{fig:PathExp}). Most of songs undergoes these simple growth-and-decay except atypical examples due to external events. In the past, the \textit{Billboard Chart} and related studies \cite{Bhattacharjee2007, Bhattacharjee2007a, Bradlow2001} only used partial information in rank history to describe and predict the whole character of rise and fall. However, one may question that even the mass media has changed greatly, will the popularity dynamics robust on the change? Directly comparing rank history in different times suggest there are significant changes in the rank patterns. In early times (Figure~\ref{fig:PopMap} a, b), songs have regular, symmetric rise and fall pattern and lifetime. Compared to earlier, in later times (Figure~\ref{fig:PopMap} c), songs have significantly slower decay while having similar rises. Most of the songs perish in few weeks, and few songs possess majority of the popularity for long time. Detailed observation in each rank history gives more pictures of this changes.

\begin{figure*}
	\centering
	\includegraphics[width=17.8cm]{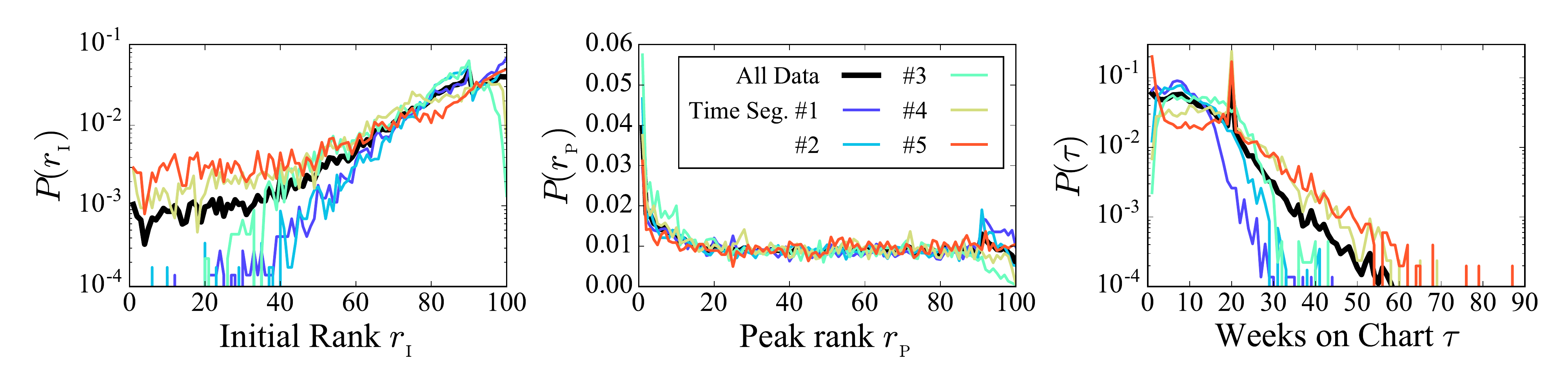}
	\caption{\label{fig:WhlStats} Rank path parameter distribution for time segments 1-5 (red-dark blue) and all times (thick black line). Distribution of initial rank $r_{_{\rm I}}$, peak rank $r_{_{\rm P}}$, final rank $r_{_{\rm I}}$, weeks on chart $\tau$, strength $s$, and rank jump $dr$ is displayed on (a-c), respectively. For details of each time segment, see table~\ref{tbl:Data}.}
\end{figure*}

As in many studies of population dynamics, ``initial rank'' $r_{\!_{\rm I}}$ distributes as exponentially decaying function from the lowest rank, with slope decreasing in time (Figure~\ref{fig:WhlStats} a). Especially, later times songs often have greater initial prosperity or shorter growth timescale. 
These abrupt entrance higher than rank 20 is also termed as ``{\em Hot-Shot Debut}'' in {\em the Billboard chart}, which is well recognized by the {\em Billboard} enthusiasts that these high-rank entrance appeared later than 1990's, and suspected to be the consequences that the nature of the popularity has been changed.

The most dramatic change happened in the survival time of a song, ``weeks on chart'' $\tau$ (Figure~\ref{fig:WhlStats} b). Survival probability in long lifetime limit decreases exponentially as in many survival analysis literature, its slope differs in times. Compared to earlier times, later times has broader tail distribution which allows songs with extraordinary lifetime, even with recurrent policy more than 50 weeks of survival. On the other hand, songs with short lifetime ($\tau<4$) behaves different by times, which is likely to influenced by chart score methods. 

Nonetheless, not all characteristics dynamically evolve in time, as seen in ``peak rank'' $r_{\!_{\rm P}}$ (Figure~\ref{fig:WhlStats} c) showing how maximum popularity distributes. Counterintuitively, songs have more possibility to have higher rank. Different from $r_{\!_{\rm I}}$ which denotes songs likely to have popularity limit regardless of the initial position.

The studies regarding rank movement have combined the features provided by ``chart numbers $r_{\!_{\rm I}}, \tau$ and $r_{\!_{\rm P}}$'' , and understand the dynamics' evolution by rescaling of the time due to spreading speed changes, \textit{i.e.} development of the mass media. Notwithstanding both early and late eras have the same exponential tail, rescaling time to collapse rank paths into universal curve fails, since times should rescaled less (faster) to match $P(\tau)$ but simultaneously rescaled more (slower) to have consistency with $P(r_{\!_{\rm I}})$. Therefore, the evolution cannot be explained by simple speed-up, but there is implicit mechanism that induces polarization of popularity between songs.

\begin{figure}
 \centering
   \includegraphics[width=\columnwidth]{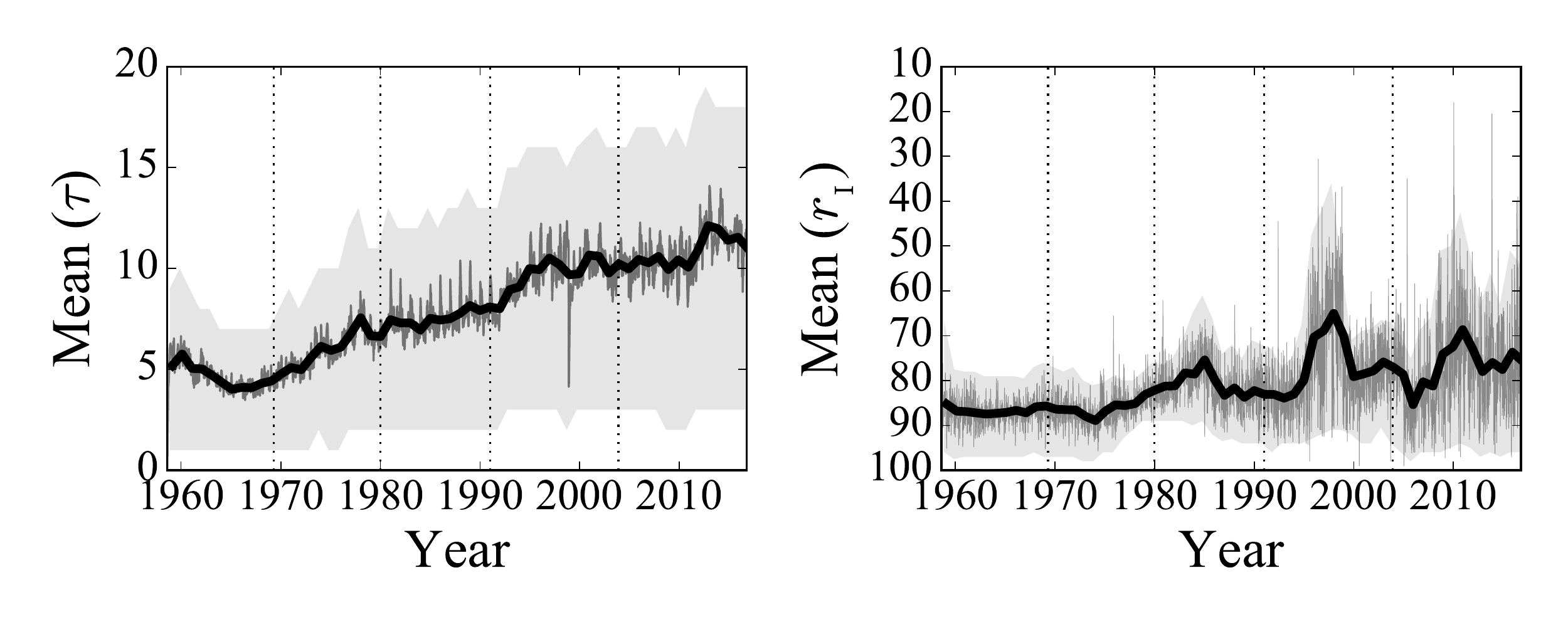}
	\caption{\label{fig:Temporal} Time dependence of $\tau$ (left) and $r_{\!_{\rm I}}$ (right) statistics. The figure shows average values for each week (thin lines) and year (thick lines). The range of upper to lower 20 percent limit is shaded. Each time segment is split by vertical lines.}
\end{figure}

Averaged values for time shows more detailed evolution of $\tau$ and $r_{\!_{\rm I}}$ (Fig.~\ref{fig:Temporal}). Average lifetime increase significantly, however its contribution is due to long-living songs than global increase of lifetimes, namely diversification of the lifetime.

Some of the rank scoring policy makes not only global shifts, but also creates sharp discontinuity of each distribution, especially the discontinuity in $P(\tau)$. This is due to recurrent policy (see~\ref{sec:Dat}) that immediately removes the single at rank under 50 at $\tau=20$. Several others including $r_{_{\rm I}}, r_{_{\rm P}} > 90$ are also suspected to be originated from unstated ranking policy. In further analyses, we limited our rank scope to 50 to avoid these irregularities.

\begin{figure*}[t]
 \centering
   \includegraphics[width=\textwidth]{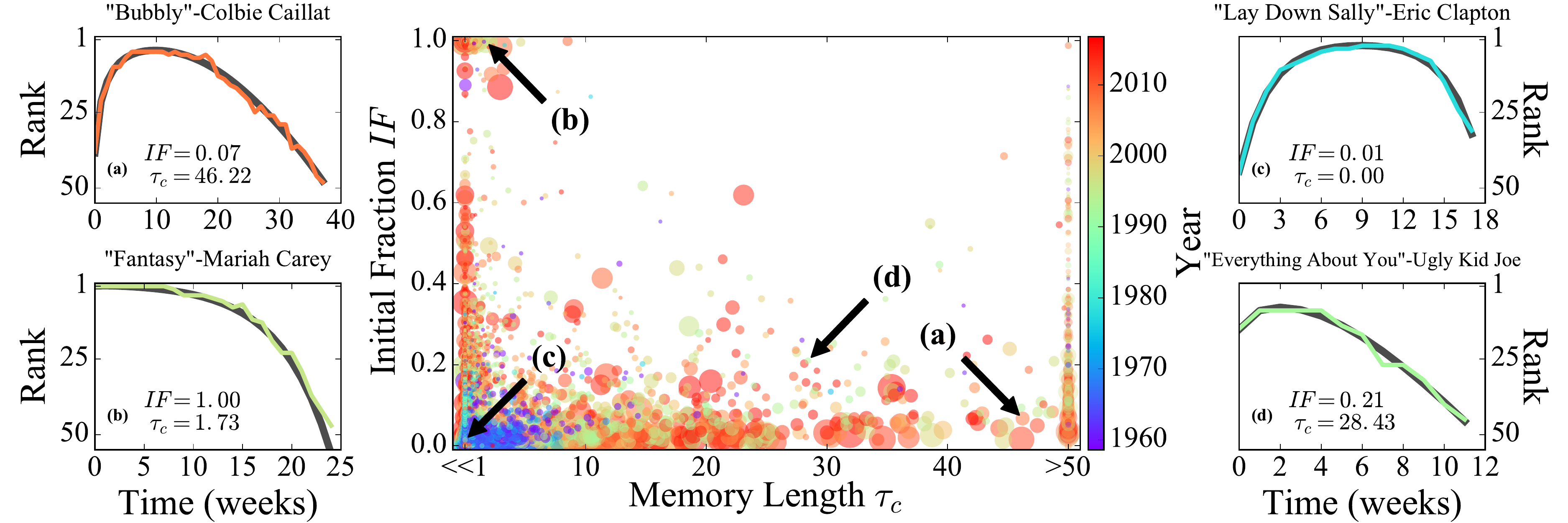}
	\caption{\label{fig:Fitter} Scatter plot of songs into $\tau_c - IF$ plane. All songs with lifetime on $r<50$ for more than 10 weeks are displayed. Size and color of points indicate strength and time of debut. In Earlier times (dark blue to green), songs are mostly have similar shape, regardless of the popularity. Ranks and optimal model estimation are shown in separate figures for four specific cases, whose $(\tau_c, IF)$ is indicated in the main figure. (a-d, see main text for details of each songs.)}
\end{figure*}


To reveal underlining rank dynamics, Zipf's law may use as translation between ranks to popularity. We translated billboard rank to popularity with assumption that the popularity follows Zipf's law~\cite{Newman2005, Clauset2009}. Zipf's law has a virtue that regardless of the entries, popularity of song with rank $r$, distributes as power law as $P(r) \sim r^{-1}$, which is likely to be hold in high-end popularity limit. Since the billboard chart enlists most popular music in whole population, popularity of high-ranked songs is directly estimated. 

We assumed that the song's activity $S(t)$ is based on spreading process and popularity $V(t)$ is sum of total activity at each time weighted by memory function as
\begin{equation}
V(t) = \int^{t}_{0} S(t-s) \Theta(s) ds.
\end{equation}
Where spreading process $S(t)$ is a nonnegative function describes the popularity from newly introduced individuals. The simplest spreading process is described by logistic growth process, which is ideal system that have finite capacity and memoryless. The equation of process is written in terms of amplitude $A$, time scale $t_0$ and initial fraction $IF$,
\begin{equation}
\label{eq:St}
S(t) = \frac{d}{dt}\frac{A}{1+(1/IF-1)e^{-t/t_0}}.
\end{equation}

Memory kernel $\Theta(s)$ describes activity response function for spreading event. In many physical system, the response is instant ($\Theta(s) = \delta(s)$) or have short time scales ($\Theta(s) \sim exp(-s/s_0)$). On the other hand, systems in critical phase, glassy, nonequilibrium or social systems often exhibit long-term memory of form $\Theta(s) \sim s^{-1+\theta}$. Previous studies report memory exponent $\theta \neq 0$ depending on the characteristics of the spreading system~\cite{Sornette2004}, however we fixed $\theta = 0$ to reduce number of free variables. 

In our claim of varying long-term memory effect along time, we added cutoff time $\tau_c$ to memory kernel $\Theta(s)$ as
\begin{equation}
\Theta(s) = A' s^{-1}e^{-s/\tau_c}.
\end{equation}

Combining two components together, we write equation of whole process in integral equation with discrete time, 
\begin{equation}
\label{eq:Vt}
V(t) = \sum^{t}_{s=0} S(t-s) (s+1)^{-1}e^{-s/\tau_c}.
\end{equation}

Figure~\ref{fig:Fitter} shows every song in $IF-\tau_c$ space, with four specific examples of different regimes. Long $\tau_c$ and small $IF$ song (figure \ref{fig:Fitter} a) have large and fast rise but falls slowly. This is in contrast to similarly short $IF$ but short $\tau_c$ case (figure \ref{fig:Fitter} c), having similar rise pattern but falls fast. On the other hand, as $IF$ increases songs experiences shorter rise, as shown on figure \ref{fig:Fitter} d.  Then, when $IF > 0.5$ songs rank only falls regardless of the $\tau_c$ as figure \ref{fig:Fitter} b. In general, for early times (dark blue to green points) large fraction of the songs has similar rank evolution; starting from small $IF$ and $\tau_c$. This is compared to later times (green to red points) where songs have diverse rank evolution as shown in scattered points. 

\begin{figure*}[t]
 \centering
   \includegraphics[width=11.4cm]{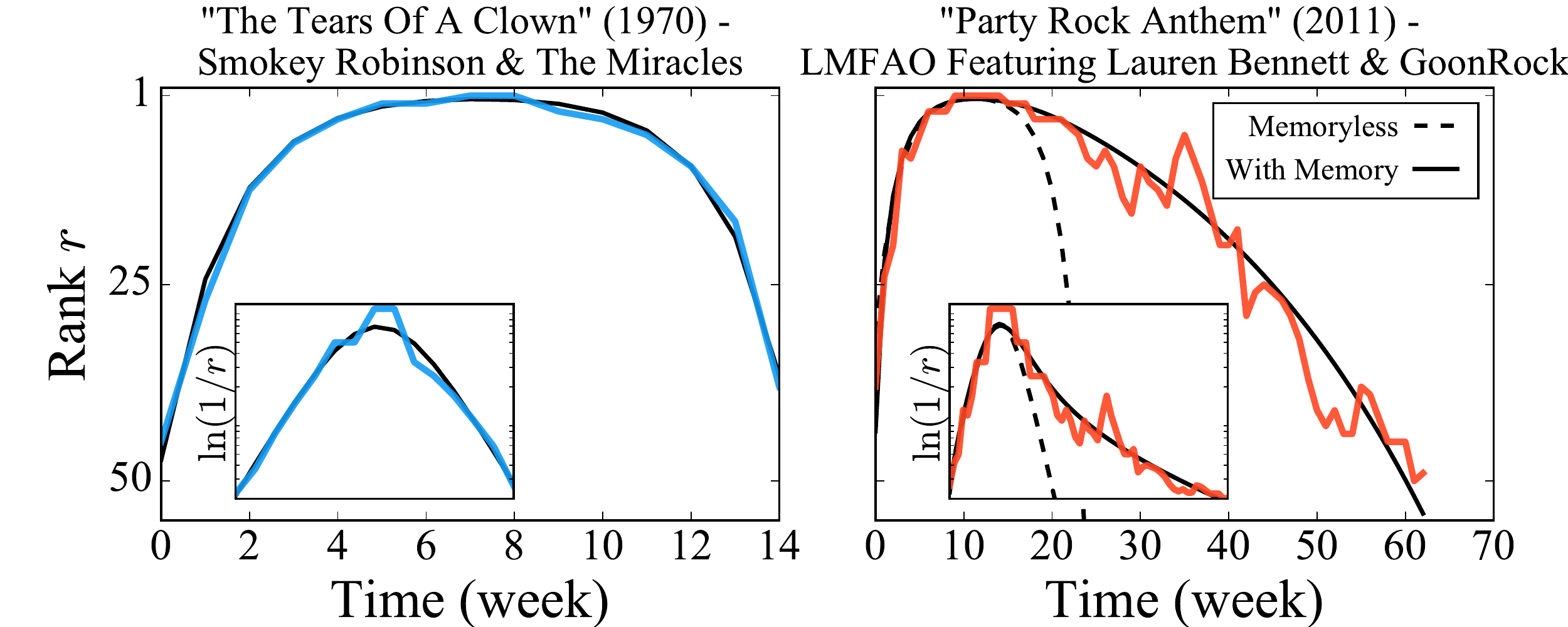}
	\caption{\label{fig:ShortAndLong} Rank movements and optimal model estimation for song with (right) and without (left) memory. Insets show the same plot for $\ln(1/r)$, which is closely related to extensive popularity. Dashed line shows estimation for the same rise without long-term memory.}
\end{figure*}

Long-term memory effect plays an important role in decay pattern as seen in figure~\ref{fig:ShortAndLong}, showing difference between rank movement with and without memory. While both rank movements with and without memory has decaying form, the difference is visible in popularity-wise plot. (Figure~\ref{fig:ShortAndLong} inset) Song without memory well resembles logistic growth characterized by exponential growth and decay with the same rate (Eq.~\ref{eq:St}), however song with long-term memory has slower decay compared to memoryless function.

\subsection*{Emergence of Long-term Memory}

\begin{figure*}[t]
 \centering
   \includegraphics[width=11.4cm]{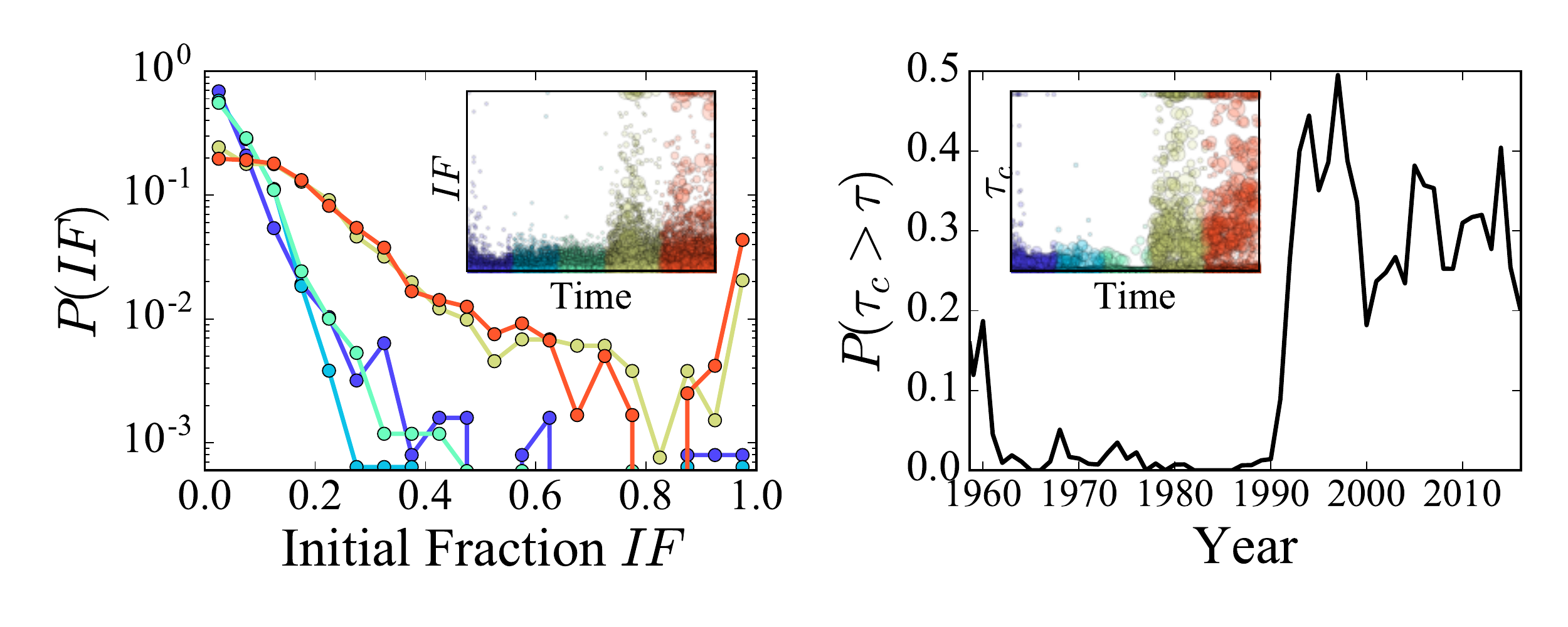}
	\caption{\label{fig:FittingYearlyStats} $IF$ and $\tau_c$ distribution for time. (Left) $IF$ distribution is shown for each time segments. In early times (dark blue to light green lines), most of songs have low $IF$s while late times (yellow and red lines) have broader distribution with finite fraction of peak starts ($IF>0.95$). (Right) Fraction of songs with memory ($\tau_c > 5$) for each year is plotted. Insets in both panels show every songs by time.}
\end{figure*}

We now look into the distribution of $IF$ and $\tau_c$ for each time segments. There are differences at certain time that explicit statistics such as $\tau$ or $r_{\!_{I}}$ showing evolution of popularity dynamics, however it is difficult to pinpoint the time of change by those values as seen in figure~\ref{fig:Temporal}. However, our model parameters show drastic difference between before and after time segment 4. (Figure~\ref{fig:FittingYearlyStats}) In early times $IF$s distributes exponentially, while later times have broader distribution. Likewise, songs that shows ``apparent'' memory effect (e.g., $\tau_c>5$) also have abrupt jump at the same point, and later times have finite fraction of the songs with memory effect.

The emergence of large $IF$ and $\tau_c$ is largely influenced by popularity factors. Large $IF$s are due to abnormally large initial ``fans'' that suppresses endogenous growth. This initial fans are from emergence of new popularity factors that are capable of attract large popularity initially. Similar argument holds for $\tau_c$, that new popularity factor has affected the memory nature of the popularity. Still, not all songs are under influences of long-memory even in latest years, since some part of the rank score may reflect long-term popularity and the others are not, and they are reflected in the rank by their fraction.

\section{Discussion}
Ranking and its dynamics provides core information on popularity of various items. In our work, we investigated such popularity dominated by preference of individuals, the newly born music. The dataset of our interest was {\em the Billboard} chart, as a representative of music ranking charts. Traditional studies in popularity dynamics less focused on the dynamics' long-term evolution, where in most cases data's time range does not span sufficiently long time, or the dynamic itself does not vary in time. Our study focused to timewise evolution of popularity dynamics through whole set of data which spans several decades, which data spans long time range enough to observe meaningful change.

Timewise statistics of singles showed that the ranking dynamics has evolved itself over decades; both lifetime and high ranks are focused to few number of songs as time goes. This inequality of song popularity rises especially after 1990's, where music distribution and consumption had drastic change from technology evolution. To catch ranking dynamics characters, we proposed epidemic type model with long-term memory. By tuning memory and epidemic parameters, we obtained popularity picture which spans throughout whole dataset. As a result, we found that popularity dynamics in recent times are more influenced by long memory. 

The origin of long-term memory is mainly due to distribution and consumption method; however the data type is another factor which should not be ignored. After 1992, {\em the Billboard Hot 100} chart included radio play into a ranking score, as well as online streaming later. Traditional scores based on sales-based score (\textit{e.g.} album, singles, online sales) are close to increment of the popularity, while activity-based score (\textit{e.g.} radio/broadcast streaming, online play, YouTube streaming) reflects total popularity by whole population and implies long memory effect. The ranking policy stands between two limits; news-like (sales-based) and history-like (activity-based), and the optimum policy is determined by character of \textit{the Billboard} magazine.

It is noteworthy to argue that throughout this study we have assumed several important statements. Firstly, we assumed that the chart ranking reflects popularity of songs but does not affect the popularity itself. However, except few cases which does not persuade consumer's preference, most of ranking statistics itself also acts as popularity pressure toward the system; in other words, ranking chart interacts with system's popularity. This interaction opens a possibility of heavy-tailed popularity distribution from Yule-Simon type process or preferential attachment which our model implies in a roundabout way by introducing long-term memory. Another fundamental assumption is that every song is independent of each other, so that popularity of new song does not have initial ``fans''. Nonetheless, our data shows there is nontrivial influences of the songs that are from same artist, which its behavior is shown by two rise-fall peaks violating our simple growth and decay pattern. Also, exogenous shocks which is shown as long re-entrance, are out of our consideration in this study. These pattern due to external events are worth investigating. Finally, we would like to emphasize that our approach to the ranking chart dynamics is applicable for any kind of charts. It will be interesting to compare ranking structure and dynamics evolution for music and other ranking in all over the world, and relate to memory and spreading characteristics of the society of its constituents.

\section{Materials and Methods}
\subsection*{The {\em Billboard} chart database \label{sec:Dat}}
We collected all songs' rank history in the {\em Billboard} Hot 100 chart since the first volume in May 8, 1958 to Oct 8, 2016, 3036 weeks in total, including 28312 songs of 8000 artists are enlisted. Data was electronically collected in the {\em Billboard.com} online chart. Song with the same title and artist is considered as the same one even if they were not enlisted in the chart in successive weeks, however arrangements, remix, or remakes were considered as different listing. Chart re-entrant of more than 3 weeks are considered as different songs. {\em Recurrents} policy removes singles which has fallen below rank 50 after 20 weeks from the chart, since 1991~\cite{BBChartLegend}.

Whole dataset is divided by five subsets with important events that affects the rankings as well as each time segment to have time span. Ranking policy changes, (Airplay and single sales by Nielsen (1991), Recurrents (1991), and Online Streaming and paid digital downloads (2005)) as well as advances in media (First widespread cassette player (1979) and portable MP3 player (1998)). Data statistics are summarized in table~\ref{tbl:Data}.

\begin{table}[]
\caption{\label{tbl:Data} Data statistics of each data subset.}
\begin{tabular}{c|r@{-}l|c|c}
\hline
\begin{tabular}{@{}c@{}}Segment \\ Number\end{tabular} & \multicolumn{2}{|c|}{Duration} & \begin{tabular}{@{}c@{}}Number of \\ Singles\end{tabular}&\begin{tabular}{@{}c@{}}Number of \\ Singers\end{tabular} \\
\hline\hline
1 & Aug \enspace 9, 1958\enspace & \enspace Apr 12, 1969 & 7,208 & 2086\\
\hline
2 & Apr 19, 1969\enspace & \enspace Dec 22, 1979 & 5,763 & 1,932\\
\hline
3 &  Dec 29, 1979\enspace & \enspace Dec 29, 1990 & 4,495 & 1,647\\	 	 	
\hline
4 & Jan \enspace5, 1991\enspace & \enspace Nov 22, 2003 & 4,238 & 2,093 \\
\hline
5 & Nov 29, 2003\enspace & \enspace Oct \enspace8, 2016 & 5,032 & 2,392 \\
\hline\hline
All time & Aug \enspace 9, 1958\enspace & \enspace Oct \enspace 8, 2016 & 26,836& 8,989 \\
\hline
\end{tabular}
\end{table}

\subsection*{Parameter estimation}
\begin{figure*}[bt]
 \centering
   \includegraphics[width=\columnwidth]{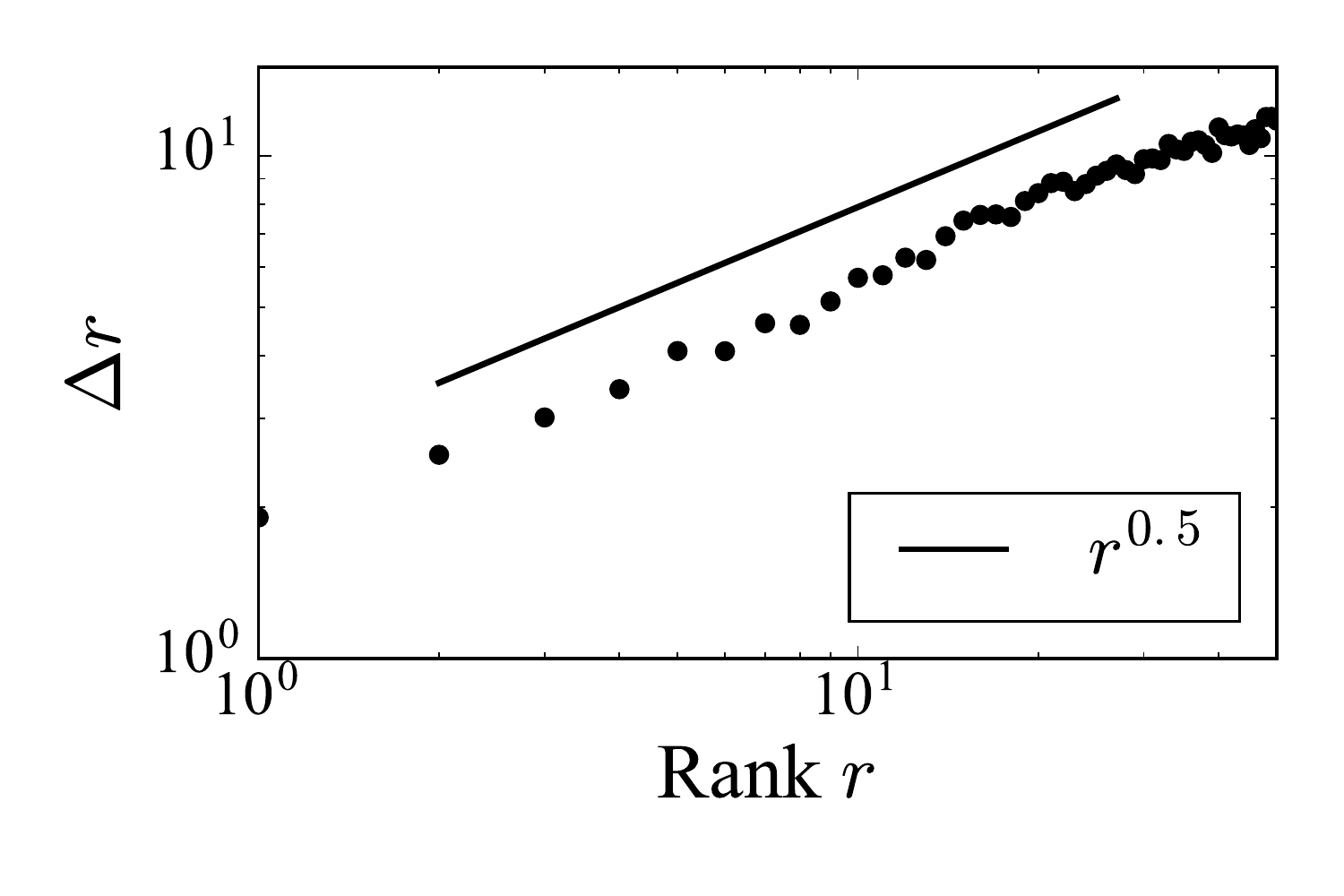}
	\caption{\label{fig:RankByVar} Fluctuation of rank. For songs in higher rank, jump size becomes narrower as shown in increasing function of rank. The line $r^{1/2}$ are guides to the eye. }
\end{figure*}

For a given rank history, equation (Eq.~\ref{eq:Vt}) is used as basis to estimate parameters $A, t_0, IF, \tau_c$ with least square optimization with trust region reflective (TRF) method. Only ranks between 1 to 50 is used to avoid inconsistency from recurrent policy. Due to different error tolerance per rank, least squares are weighted by rank with weights $(\Delta r) \sim r^{1/2}$, using observation from rank deviation (see Fig.~\ref{fig:RankByVar}). 


\section*{Acknowledgments}
This research was supported by Basic Science Research Program through the National Research Foundation of Korea (NRF) by the Korea government [NRF-2017R1A2B3006930 (H.S., H.J.)].


\bibliography{Billboard-References}

\begin{figure*}[b]
 \centering
   \includegraphics[width=17.8cm]{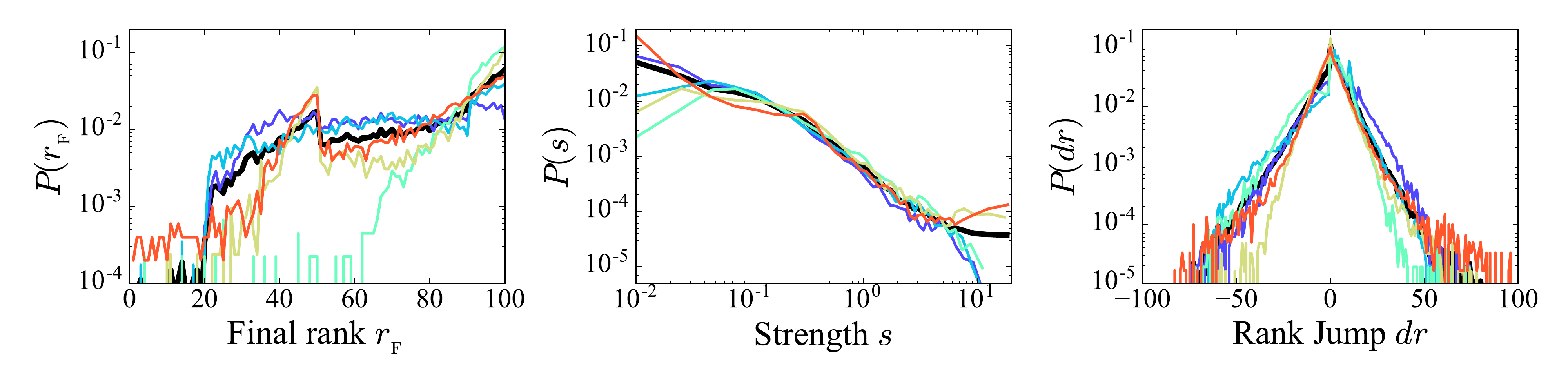}
	\caption{\label{fig:WhlStats-sup} Frequency of final rank $r_{\!_{\rm F}}$, strength $s$, and rank jump $dr$ is displayed on (a-c) respectively. Each frequency is displayed for time segments 1 to 5 (red to violet) and all times (black). (a) Frequency of rank at the last week of songs (Final rank $r_{\!_{\rm F}}$). In early times (time segment 1, 2), a song exits chart at almost equal ranks below $r_{\!_{\rm F}}>30$ but rapidly decreases above $r_{\!_{\rm F}}<30$, while in later times (time segment 4, 5), lower rank is more likely to exit as seen increasing slope by $r_{\!_{\rm F}}$ except special jump at $r_{\!_{\rm F}}=50$ due to recurrents. It seems more probable to have less popular songs likely to exit the chart as in later era, however final rank distribution have counterintuitive pattern. Similar pattern is repeated in final rank $r_{\!_{\rm F}}$ (Fig~\ref{fig:WhlStats} b) as frequency of chart exit at higher rank is exponentially decreasing. (b) Frequency of $s = \sum_t 1/r_t$ (strength) of songs. (c) Rank jump $dr$ is rank difference between successive weeks. rank transition probability, which presents exponential decay in both rise and fall with different jump scale (slope of the line). In average, rising (falling) is faster (slower) than each other, which gives typical pattern of a path; rises slowly, and falls rapidly. This picture may counterintuitive to typical famous songs that gains popularity abruptly and falls gradually. For details of each time segment, see table~\ref{tbl:Data}.}
\end{figure*}

\begin{figure*}[bt]
 \centering
   \includegraphics[width=11.4cm]{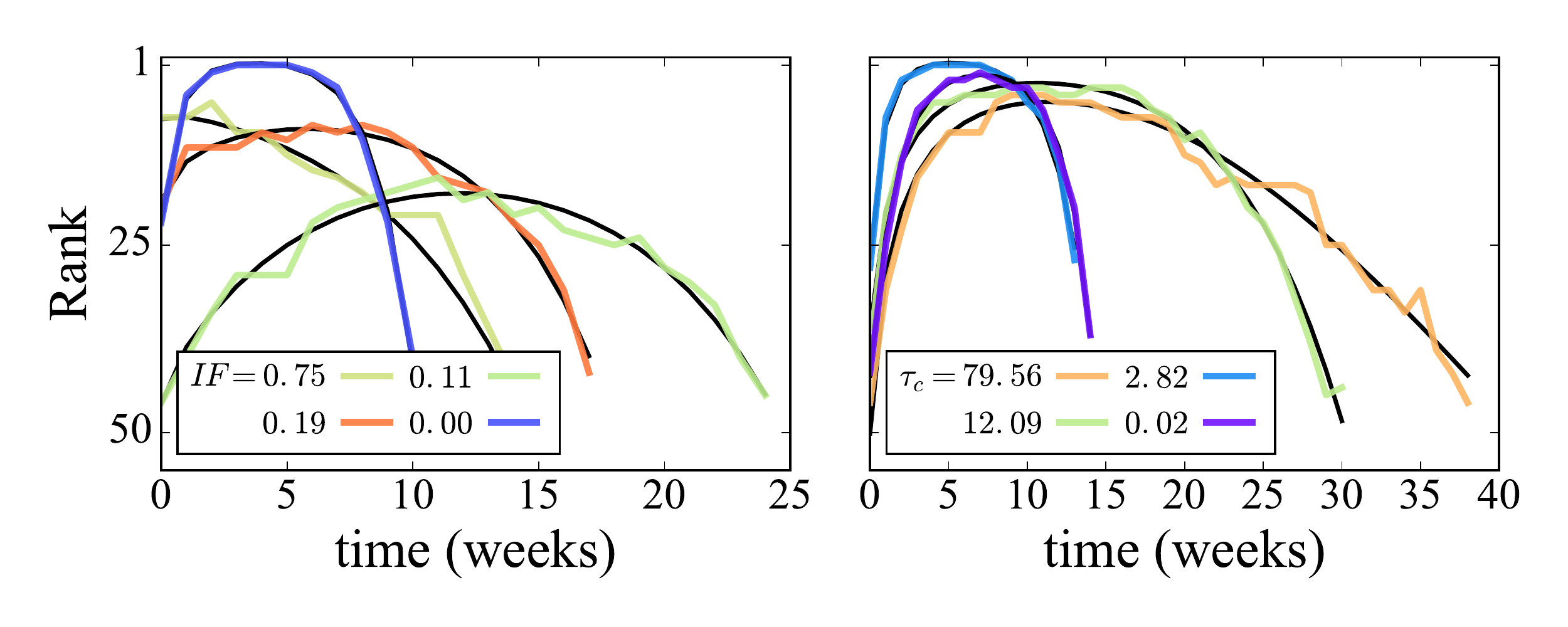}
	\caption{\label{fig:ParamChars} Comparison of songs with varying $IF$ and $\tau_c$. (Left) Four songs with $\tau_c < 1$ are shown,``{\em This Is For The Lover In You}''(1997) -- Babyface Feat. LL Cool J, Howard Hewett, Jody Watley \& Jef ($IF=0.75$), ``{\em Shawty}''(2007) -- Plies Featuring T-Pain ($IF=0.19$), ``{\em I'll Stand By You}''(1995) -- Pretenders ($IF=0.11$), ``{\em I Feel Fine}''(1965) -- The Beatles ($IF=0.00$). $IF$ is indicates starting point of the growth, therefore $IF$ increases, difference between initial rank to peak rank decreases. Without memory, $IF$ is equal to fraction of initial to accumulated popularity. Small $IF$ means initial rank is small compared to peak rank, and rank only decreases if $IF>0.5$ therefore in this case initial rank is equal to peak rank. (Right) Four songs with $IF<0.01$ are shown, ``{\em A Thousand Miles}''(2002) -- Vanessa Carlton ($\tau_c = 79.56$), ``{\em Always}''(1995) -- Bon Jovi ($\tau_c = 12.09$), ``{\em Honky Tonk Women}''(1969) -- The Rolling Stones ($\tau_c = 2.82$), ``{\em Handy Man}''(1960) -- Jimmy Jones ($\tau_c = 0.02$). As $\tau_c$ increases, songs have slower rank decay. $\tau_c$ governs memory strength, which characterizes slowdown of the popularity decay. When $\tau_c>1$, past popularity has effective contribution to following weeks, leads visible asymmetry between rise and fall. For $\tau_c \sim \tau$ songs long-term memory is present in its overall rank dynamics.}
\end{figure*}

\begin{figure*}[bt]
 \centering
   \includegraphics[width=11.4cm]{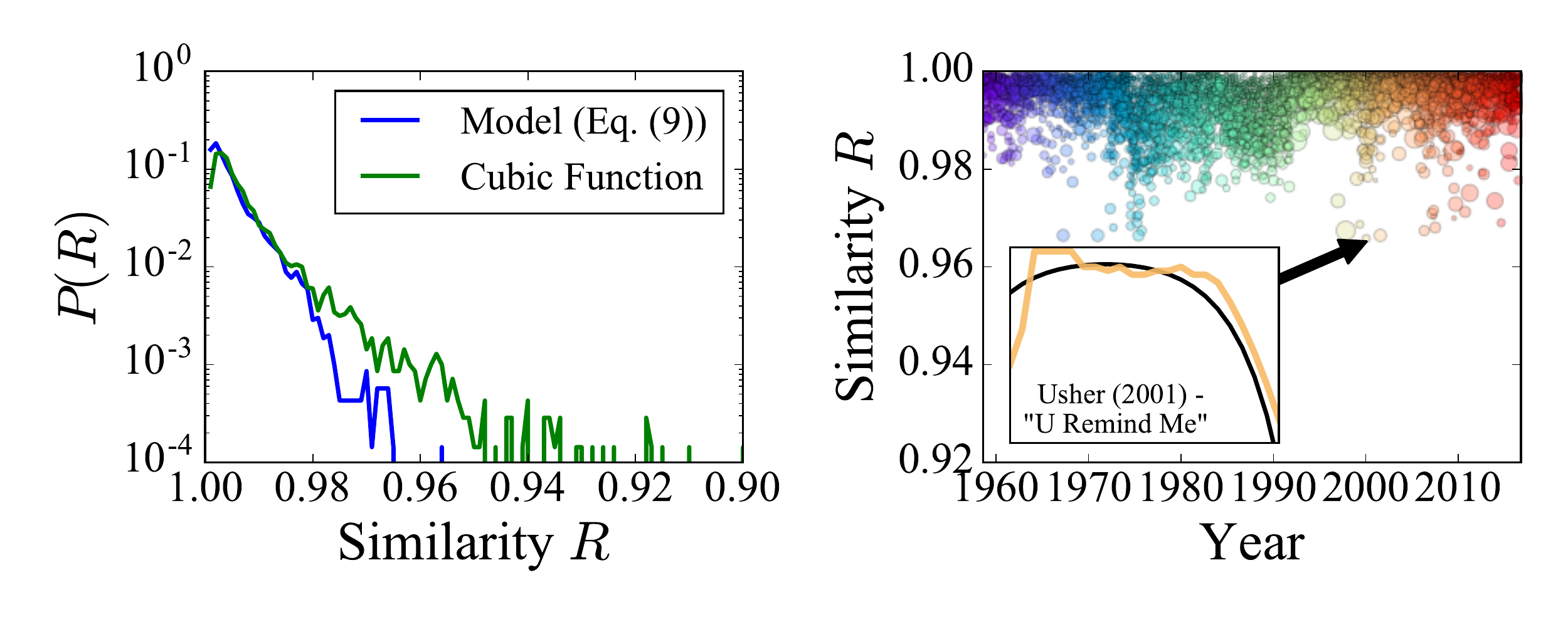}
	\caption{\label{fig:SimilarityStats} Statistics of rank similarity $R$ between optimal model estimation $y_t$ and data $r_t$, defined as $R[r_t, y_t] = {\sum_1^\tau{r_t y_t}}/{\sqrt{\sum_1^\tau {r^2_t} \sum_1^\tau {y^2_t}}}$. (Left) Distribution of similarity $R$. The model's accuracy was compared to null model with the same degrees of freedom $(y'_t = a+bt+ct^2+dt^3)$, whose distribution indicates our model has better estimation. denotes our model has apparently catches global trend that the naive fitting without knowing any functional form of rank sequence cannot have. (Right) The similarity for each songs are plotted by time. The inset shows a song with one of the most deviating estimation, which rank movement behaves different to our model due to external factors. Our model estimates regardless of the time, notwithstanding that the rank dynamics has dramatic evolution over time.
	}
\end{figure*}

\end{document}